\documentclass[apj]{emulateapj}
\usepackage{apjfonts}
\slugcomment{Accepted by ApJ Letters}
\shortauthors{Tran et al.}
\shorttitle{Late Assembly of Massive Cluster Galaxies}

\begin{document}

\newcommand{\kms}{~km~s$^{-1}$}
\newcommand{\logh}{$+5\log h$}
\newcommand{\mipsmu}{$24\mu$m}

\title{The Late Stellar Assembly of Massive Cluster Galaxies Via Major Merging}

\author{Kim-Vy H. Tran$^1$, John Moustakas$^2$, Anthony
  H. Gonzalez$^3$, Lei Bai$^4$, Dennis Zaritsky$^4$, \& Stefan
  J. Kautsch$^3$}

\footnotetext[1]{Institute for Theoretical Physics, University of
  Z\"urich, CH-8057 Z\"urich, Switzerland}
\footnotetext[2]{Center for Cosmology and Particle Physics, 4
Washington Place, New York University, New York, NY 10003}
\footnotetext[3]{Department of Astronomy, University of Florida,
  Gainesville, FL 32611}
\footnotetext[4]{Steward Observatory, University of Arizona, 933 
North Cherry Avenue, Tucson, AZ 85721}

\begin{abstract}

We present multi-wavelength observations of the brightest galaxies in
four X-ray luminous groups at $z\sim0.37$ that will merge to form a
cluster comparable in mass to Coma.  Ordered by increasing stellar
mass, the four brightest group galaxies (BGGs) present a time sequence
where BGG-1, 2, \& 3 are in merging systems and BGG-4 is a massive
remnant ($M_{\ast}=6.7\times10^{11}M_{\odot}$).  BGG-1 \& 2 have
bright, gravitationally bound companions and BGG-3 has two nuclei
separated by only 2.5~kpc, thus merging at $z<0.5$ increases the
BGG mass by $\gtrsim40$\% ($t_{MGR}<2$ Gyr) and $V$-band luminosity by
$\sim0.4$ mag.  The BGGs' rest-frame $(B-V)$ colors correspond to
stellar ages of $>3$ Gyr, and their tight scatter in $(B-V)$ color
($\sigma_{BV}=0.032$) confirms they formed the bulk of their stars at
$z>0.9$.  Optical spectroscopy shows no signs of recent ($<1.5$ Gyr)
or ongoing star formation.  Only two BGGs are weakly detected at
\mipsmu, and X-ray and optical data indicate the emission in BGG-2 is
due to an AGN.  All four BGGs and their companions are early-type
(bulge-dominated) galaxies, and they are embedded in diffuse stellar
envelopes up to $\sim140$~kpc across.  The four BGG systems must
evolve into the massive, red, early-type galaxies dominating local
clusters.  Our results show that: 1) massive galaxies in groups and
clusters form via dissipationless merging; and 2) the group
environment is critical for this process.

\end{abstract}


\section{Introduction}

Whether massive cluster galaxies at $z\sim0$ build up a significant
fraction of their stellar masses at $z<1$ via merging continues to be
an intensely debated question.  In the paradigm of hierarchical
formation, galaxies continue to grow via accretion of smaller
satellites \citep{peebles:70}, and recent models predict that
brightest cluster galaxies assemble half their stellar masses at
$z<0.5$ \citep{delucia:07b}.  However, direct merging is not favored
in virialized clusters due to their high velocity dispersions
($\sigma_{1D}\sim1000$~\kms).

Brightest cluster galaxies (BCG) are particularly compelling tracers
of mass growth because they are the most massive galaxies in the
universe, can be observed to high redshifts \citep{gunn:75}, and
should have a rich merger history.  A number of BCGs in the local
universe have multiple nuclei \citep{hoessel:85,lauer:86,laine:03}
which suggests recent merging.  However, studies of BCGs in the more
distant universe disagree as to whether BCGs have grown significantly
in stellar mass since redshift $z\sim1$ \citep{brough:02,whiley:08}.

While merging of red, luminous members is observed in groups and
dynamically young clusters \citep{vandokkum:99,tran:05b,mcintosh:07},
the merging frequency and thus its importance remains uncertain.  The
key may be environment: merging is rare in established clusters but
more frequent in groups that are accreted onto clusters. Because
groups have a higher fraction of passive early-type galaxies relative
to the field \citep{wilman:05,jeltema:07}, merging between such
systems will form massive, passive galaxies.

Our study focuses on the brightest galaxies in four X-ray luminous
galaxy groups at $z\sim0.37$ discovered in the {\it Las Campanas
Distant Cluster Survey}\footnote{The LCDCS was designed to find galaxy
clusters at $z\sim0.35-0.9$ and was sensitive to groups only at the
lowest redshifts.} \citep{gonzalez:01}.  These four galaxy groups are
gravitationally bound to each other and form an extended structure
that will assemble into a galaxy cluster comparable in mass to Coma by
the present epoch \citep[hereafter G05]{gonzalez:05}.  By studying
galaxies in the group environment prior to cluster assembly, we
directly test whether galaxy-galaxy merging on group scales can drive
stellar growth in the massive galaxies observed in local clusters, and
whether a large fraction of new stars is formed in the process.
Throughout the paper, we use $H_0=70$~km~s$^{-1}$~Mpc$^{-1}$,
$\Omega_M=0.3$, and $\Omega_{\Lambda}=0.7$; at $z=0.37$, this
corresponds to a scale of 5.12 kpc per arcsec and a lookback time of 4
Gyr.

\section{Observations}

Follow-up $Chandra$ ACIS-I imaging and optical spectroscopy of the
unique super-group 1120 field confirmed the existence of four X-ray
luminous groups at $z\sim0.37$ (hereafter SG1120).  These groups lie
within a projected region of $\sim3$~Mpc and have X-ray temperatures
of $T_X=1.7-3.0$ keV (G05).  Our study combines imaging and
spectroscopy from six different observatories\footnote{This work is
based on observations made with 1) the NASA/ESA {\it Hubble Space
Telescope}, obtained at the Space Telescope Science Institute, which
is operated by AURA, Inc., under NASA contract NAS 5-26555; 2) the
{\it Spitzer Space Telescope}, which is operated by the Jet Propulsion
Laboratory, California Institute of Technology under a contract with
NASA; 3) the {\it Chandra X-ray Telescope}, obtained at the Chandra
X-ray Observatory Center, which is operated by the Smithsonian
Astrophysics Observatory for and on behalf of NASA under contract
NAS8-03060; 4) the 4 meter Mayall telescope at Kitt Peak National
Observatory; 5) the 6.5 meter Magellan Telescopes at Las Campanas
Observatory, Chile; and 6) the ESO Telescopes at the Paranal
Observatories (072.A-0367, 076.B-0362, and 078.B-0409).}.


\subsection{Optical \& Near-Infrared Imaging}\label{kcorr}

The optical photometry is measured from VLT/VIMOS imaging in $BVR$ for
a nearly contiguous $18'\times20'$ region, and from a contiguous
mosaic ($11'\times18'$) composed of 10 pointings taken with the
HST/ACS in F814W.  The high resolution ACS imaging is critical for
identifying close galaxy pairs and merger signatures.  Two pointings
with KPNO/FLAMINGOS also provide a contiguous $K_s$ mosaic
($16'\times19'$).

Using the HST/ACS mosaic as the master detection image, line-matched
catalogs were generated using SExtractor v2.5.0 \citep{bertin:96}.
Rest-frame absolute magnitudes, K-corrections, and stellar masses were
determined with {\it k-correct} v4.1 \citep{blanton:07} which uses
templates based on models from \citet[hereafter BC03]{bruzual:03}.  We
used MAG$_{-}$AUTO photometry from the $BVRK_s$ imaging and assumed
minimum photometric uncertainties in each bandpass of 0.05 mag. The
photometry have also been corrected for foreground Galactic extinction
using the \citet{schlegel:98} dust maps and the \citet{odonnell:94}
Milky Way extinction curve, assuming $R_V=3.1$.

\subsection{Optical Spectroscopy}

The initial survey using VLT/VIMOS confirmed that four of the five
X-ray regions are galaxy groups at $z\sim0.37$ while the fifth is a
cluster at $z=0.48$ (G05).  Targets were selected from a
magnitude-limited catalog ($R\leq22.5$ mag), and the spectra reduced
using a combination of IRAF\footnote{IRAF is distributed by the
National Optical Astronomy Observatories.} routines and custom
software; see \citet{tran:05a} for details on the reduction and
redshift determination.

Further spectroscopy was obtained using Magellan/LDSS3 and VLT/FORS2.
In both cases, targets were selected from $K_s$ catalogs ($K_s\leq20$
mag); for reference, the brightest supergroup galaxy has $K_s=14.62$
mag.  Guided by the redshift distribution, we define the supergroup's
redshift range to be $0.32\leq z\leq0.39$ where the brightest group
galaxies lie at $0.354\leq z\leq0.372$ (Table~\ref{tab:bgg}).  These
spectra provided 97 new members for a total of 198 confirmed
supergroup galaxies across the approximately $20'\times20'$ region.

The medium resolution spectroscopy corresponds to 0.7\AA~pix$^{-1}$
(LDSS3), 1.65\AA~pix$^{-1}$ (FORS2), and 2.5\AA~pix$^{-1}$ (VIMOS).
The spectral range for all three instruments covers [OII]$\lambda3727$
to [OIII]$\lambda5007$ for most supergroup members.

\begin{deluxetable}{lrrrrrr}
\tablecaption{Properties of Brightest Group Galaxies\tablenotemark{a}}
\tablewidth{0pt}
\tablehead{
\colhead{BGG}	&	\colhead{$z$}	&
\colhead{$V$}	&	\colhead{$(B-V)$} &
\colhead{$M_{\ast}$}	&
\colhead{$L_{IR}$\tablenotemark{b}}	&
\colhead{$\sigma_{1D}$ \tablenotemark{c}} \\
& & & & \colhead{($M_{\odot}$)} & \colhead{(erg~s$^{-1}$)} &
\colhead{(km~s$^{-1}$)} 
}
\startdata
1a & 0.3540 & $-22.46$ & 0.86 & $2.9\times10^{11}$ & \nodata & $329\pm66$ \\
1b & 0.3532 & $-21.49$ & 0.85 & $1.2\times10^{11}$ & \nodata & $329\pm66$ \\
2a & 0.3706 & $-22.97$ & 0.94 & $5.0\times10^{11}$ & $1.5\times10^{44}$ & $421\pm83$ \\
2b & 0.3704 & $-22.04$ & 0.92 & $2.1\times10^{11}$ & \nodata & $421\pm83$ \\
3 & 0.3713 & $-23.20$ & 0.94 & $5.4\times10^{11}$ & $8.0\times10^{43}$ & $588\pm90$ \\
4 & 0.3720 & $-23.38$ & 0.91 & $6.7\times10^{11}$ & \nodata & $565\pm130$ \\
\enddata
\tablenotetext{a}{Positions for the individual galaxy groups are in G05.}
\tablenotetext{b}{Infrared luminosities determined from SST/MIPS 24$\mu$m
fluxes.}   
\tablenotetext{c}{Line-of-sight velocity dispersions ($\sigma_{1D}$)
are measured using all members within $500$~kpc of the BGGs (13-36
galaxies for each group) and using the biweight estimator
\citep{beers:90}. }
\label{tab:bgg}
\end{deluxetable}

\subsection{\mipsmu~Imaging}

Wide-field MIPS imaging ($0.36^{\circ}\times1^{\circ}$) was taken in
slow scan mode, and the data processed with the MIPS Data Analysis
Tool \citep[DAT version 3.02;][]{gordon:05} where array-averaged
background subtraction was applied to improve the flat-fielding.  At
\mipsmu, the exposure time is $\sim1000$~s~pix$^{-1}$ and the
3$\sigma$ point source detection limit $\sim100\mu$Jy.  We use DAOPHOT
II \citep{stetson:87} to detect sources and measure their fluxes, and
we follow the same strategy as described in \citet{papovich:04}.  The
spectroscopically confirmed members are correlated with the
\mipsmu~catalog using a $2''$ search radius.  Total IR luminosities
are determined using star-forming galaxy templates developed by
\citet{dale:02}, and $L_{IR}$ converted to star formation rates using
Eq. 3 in \citet{kennicutt:98b}.

\section{Results}

\subsection{Brightest Group Galaxies}

\begin{figure}
\epsscale{1.15}
\plotone{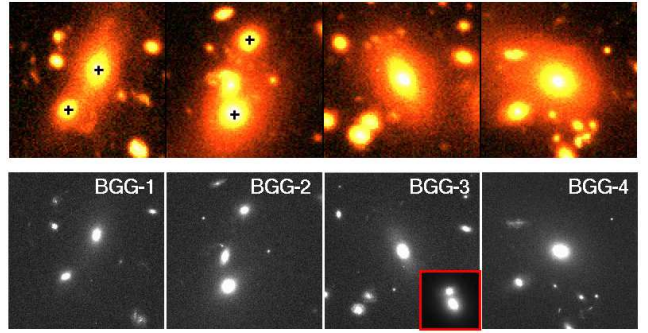}
\caption{{\it Top:} VLT/VIMOS $BVR$ images ($20''\times20''$) of
the four brightest group galaxies ordered in
increasing stellar mass, $i.e.$ a rough time sequence.  BGG-1 \& 2
both have bright, gravitationally bound companions (marked with
crosses).  All four BGGs have extended, diffuse stellar envelopes that
are up to 140~kpc across (BGG-1).  {\it Bottom:} ACS F814W imaging
of the same BGGs reveals the spectacular double nucleus in BGG-3 (image
inset; projected separation of 2.5~kpc).
\label{fig:tnails}}
\end{figure}

Images of the brightest group galaxy (BGG) within each of the four
X-ray luminous groups at $z\sim0.37$ are shown in
Fig.~\ref{fig:tnails}.  The BGGs are ordered in increasing stellar
mass (see \S\ref{kcorr}) and range from
$2.9-6.7\times10^{11}M_{\odot}$ (Table~\ref{tab:bgg}).

All of the BGGs are early-type galaxies with no visual signs of
ongoing star formation.  These BGGs are in groups with masses less
than that of galaxy clusters ($\sim10^{14}M_{\odot}$
vs. $\sim10^{15}M_{\odot}$) at $\sim1/3$ the lookback time, yet they
already morphologically resemble the massive early-type galaxies that
are common in $z\sim0$ clusters, $e.g.$ Coma \citep{davis:76}. The
BGGs are even comparable in stellar mass to the most massive brightest
{\it cluster} galaxies in the local universe
\citep[$M_{\ast}\sim5\times10^{11}M_{\odot}$;][]{vonderlinden:07}

Spectroscopy uncovers the second striking result: the two lowest mass
BGGs are part of merging systems (Fig.~\ref{fig:tnails}). BGG-1 has a
companion galaxy ($D_{proj}=34$~kpc, $\delta$v=170\kms) where both
are embedded in a common, extended envelope that has a semi-major axis
of $\sim140$~kpc; this stellar envelope is especially prominent in
the $R$-band imaging and is reminiscent of the stellar plume
associated with a major merger forming a brightest cluster galaxy at
$z=0.39$ \citep{rines:07}.  BGG-2 also has a companion galaxy
($D_{proj}=48$~kpc, $\delta$v=53\kms; note
$\delta$v$\approx\sigma_{cz}$) within a faint, extended stellar halo.

In the ground-based imaging (PSF$\sim0.7''$), BGG-3 looks like a
normal early-type galaxy but the high resolution ACS imaging ($0.05''$
pix$^{-1}$) reveals its spectacular double nucleus
(Fig.~\ref{fig:tnails}).  As part of our spectroscopic survey, we
aligned a slit along the two nuclei and confirm that both are part of
the same galaxy.  The nuclei are separated by only 2.5~kpc.  BGG-3
also has a diffuse stellar halo.

BGG-4 is the most massive galaxy in SG1120 and lies in the group with
the highest X-ray temperature (G05).  Other than a diffuse stellar
halo, BGG-4 has no obvious signs of recent ($t<1$ Gyr) merging, $e.g.$
multiple nuclei.  However, recent simulations indicate that massive
ellipticals ($M_{\ast}>6\times10^{11}M_{\odot}$) like BGG-4 form via
dissipationless merging of two spheroids \citep{naab:06}.

Taken as a whole, these four BGGs present a time sequence where
luminous ($L>L^{\ast}$) galaxies continue to merge at late times, as
predicted by hierarchical formation \citep{peebles:70}.  The merging
galaxies are spheroids that have no visible signs of current or recent
star formation, but do have extended stellar envelopes. During the
merging process, stars from these spheroids are gravitationally
disrupted and likely to contribute to the intracluster light
\citep{mihos:03}.  The resulting merger remnant will be a massive,
early-type galaxy embedded in an extended, diffuse stellar envelope,
$e.g.$ BGG-4.

\subsection{Color-Magnitude Diagram \& Stellar Mass}

\begin{figure}
\epsscale{1.0}
\plotone{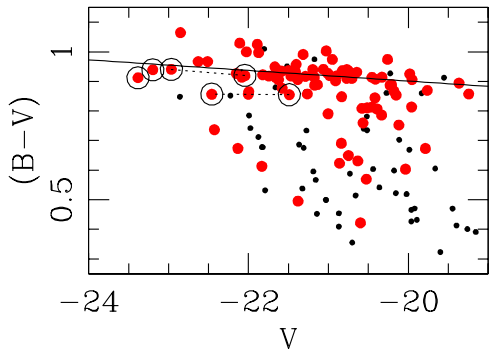}
\caption{Rest-frame color-magnitude diagram for the 146 members
($0.32\leq z\leq0.39$) on the ACS mosaic (small dots).  The large
filled dots are passive members ([OII] EW$<5$\AA) and the large open
circles denote the BGGs and merging companions (connected with the
dotted lines).  The slope of the CM relation is taken from CL~1358+62
\citep[$z=0.33$;][]{vandokkum:98a}, and the fit is normalized to the
passive members.  After merging, the two faintest BGGs will increase
their $V$-band luminosities by $\sim0.4$ mag.  
\label{fig:cmd}}
\end{figure}

Figure~\ref{fig:cmd} shows the rest-frame $(B-V)$ color versus
absolute $V$-band magnitude for the 146 supergroup members that fall
on the HST/ACS mosaic.  Passive members (N=88; $\sim60$\%), are
well-fit using the color-magnitude (CM) relation measured in
CL~1358+62, a massive galaxy cluster at $z=0.33$ \citep[hereafter
vD98]{vandokkum:98a}.  The CM relation is normalized to the passive
members where outliers have been iteratively removed using the median
absolute deviation.  We note that $\sim60$\% of the members are E/S0
galaxies.

All of the BGGs lie within $\Delta (B-V)>-0.1$ of the CM relation and
are thus red.  While they all lie blue-ward of the CM relation,
assuming the BGGs evolve passively their $(B-V)$ colors will increase
such that the BGGs will be tightly clustered around the CM relation by
$z\sim0$.  Their $(B-V)$ colors already correspond to single stellar
populations that are at least 3 Gyr old (solar metallicity; BC03).

Included in Fig.~\ref{fig:cmd} and Table~\ref{tab:bgg} are
the companion galaxies to BGG-1 and BGG-2.  Once these companions
merge with their BGGs via, $e.g.$ dynamical friction, the resulting
BGGs will be $\sim0.4$ magnitudes brighter.

The stellar masses determined from the photometry provide further
evidence of BGG build-up. The companion galaxies to BGG-1 and BGG-2
are each $\sim40$\% of their respective BGG's mass, and the two nuclei
in BGG-3 have a luminosity ratio of about 1:2.  Our data indicate that
the BGGs grow in mass by $\gtrsim40-50$\% during the most recent
merger.  We estimate the merging timescale by setting it equal to the
dynamical friction timescale \citep{binney:87} and find that
$t_{MGR}\lesssim2$ Gyr, consistent with merging timescales measured
for luminous red galaxies \citep{conroy:07} and for close pairs in
cosmological simulations \citep{kitzbichler:08}.


A powerful feature of a color-magnitude analysis is using the scatter
in color to determine the relative stellar ages of different galaxy
populations \citep{bower:98}.  The $(B-V)$ scatter (RMS) for the six
members in the BGG systems is $0.032$; this is significantly smaller
than the color scatter of the passive members ($\sigma_{BV}=0.116$;
Fig.~\ref{fig:cmd}), $i.e.$ the BGGs are more homogeneous in stellar
age than the passive population.

Because $(B-V)$ scatter decreases as stellar populations age, we also
can constrain the formation redshift of the BGGs.  For simplicity, we
model the BGGs as single stellar populations (SSPs) and allow SSPs to
form any time up to $t$ where $t_0$ is the epoch at $z=0.37$.  Using
Eq.~A3 in vD98, we estimate the $(B-V)$ scatter from 5000 SSPs
generated at random times and find that to have $\sigma_{BV}=0.032$,
all of the SSPs must form by $t=0.65t_0$.  In the current cosmology,
this corresponds to all of the stars in the BGGs forming at $z>0.9$.

Assuming the BGGs continue to evolve passively, their $(B-V)$ scatter
decreases to $\sigma_{BV}\sim0.020$ by $z=0$.  This is even smaller
than the intrinsic scatter of $\sigma_{BV}=0.028\pm0.004$ measured
using the 104 brightest $H$-band early-types in Coma at $z=0.02$
\citep{eisenhardt:07}.  Note that $\sim60-70$\% of the supergroup
galaxies are already passive and/or red (Fig.~\ref{fig:cmd}), thus
later merging events are unlikely to significantly change the massive
BGGs' mean stellar ages.

\subsection{Composite Optical Spectrum}

\begin{figure}
\epsscale{1.0}
\plotone{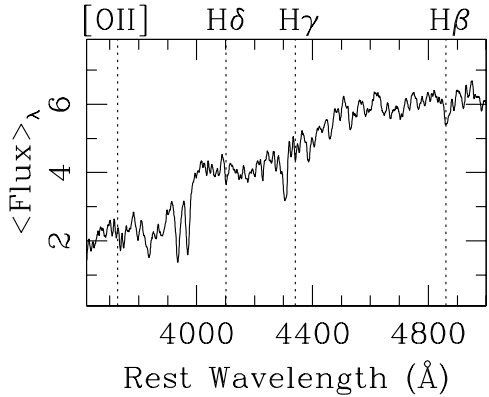}
\caption{Composite rest-frame spectrum of the six members making up the
  merging BGG systems; each spectrum was normalized to its flux level
  at $6400<\lambda<6600$\AA~before summing.  The
  high signal-to-noise spectrum shows no signs of ongoing or recent
  star formation (within $\sim1.5$ Gyr).
\label{fig:bgg1d}}
\end{figure}

Figure~\ref{fig:bgg1d} shows the composite rest-frame optical spectrum
of the six galaxies in the BGG systems.  The composite BGG spectrum
has the strong H\&K absorption features associated with an old
($>10^9$ Gyr) stellar population and has no emission features nor
Balmer absorption.  \citet{jeltema:07} also find that
intermediate-redshift ($0.2<z<0.6$) BGGs lack [OII]$\lambda3727$
emission.

Optical emission lines are physically driven by ongoing star formation
while strong Balmer absorption, H$\delta$ in particular, provides a
fossil record of any star formation that ended within the last
$\sim1.5$ Gyr \citep[$e.g.$][]{couch:87}.  The absence of these
features in the composite BGG spectrum indicates that these members
have not formed any stars recently, and the spectrum's $D_N(4000)$
index of 1.72 corresponds to a mean stellar age of $\sim1.7$ Gyr
(solar metallicity; BC03).

\subsection{\mipsmu~Detections}

Only BGG-2 \& 3 are detected at \mipsmu, and assuming the mid-infrared
emission is due to dusty star formation leads to rates of 6.7 and 3.6
$M_{\odot}$~yr$^{-1}$, respectively.  However, dusty star-forming
red-sequence galaxies detected at \mipsmu~tend to show weak
[OII]$\lambda3727$ emission in their optical spectra \citep{wolf:05},
and neither BGG-2 nor 3 do (Fig.~\ref{fig:bgg1d}).  While
\mipsmu~emission is often attributed to ongoing star formation, it can
also be due to an active galactic nucleus (AGN).  BGG-2 is an X-ray
point source, thus both its X-ray and \mipsmu~emission are likely due
to an AGN.  BGG-3's weaker \mipsmu~detection makes it difficult to
discriminate between a faint AGN and star formation.

BGG-2's multi-wavelength properties are consistent with an old stellar
population ($t_{\ast}\gtrsim2$ Gyr) and AGN.  Current models favor the
formation of luminous red galaxies via merging of progenitors with
supermassive black holes; the gravitational perturbations bring fresh
gas to the small central region and trigger an AGN phase
\citep[$e.g.$][]{springel:05a}.  BGG-2, and possibly BGG-3, are
compatible with such a model.

\section{Discussion \& Conclusions}

These four BGG snapshots provide unprecedented observational evidence
for the late ($z<0.5$) stellar assembly of massive spheroidal galaxies
in clusters via dissipationless merging of massive progenitors.  Two
of the four BGGs are part of merging pairs and another has a double
nucleus with a separation of only 2.5~kpc.  When ordered by
increasing stellar mass ($2.9-6.7\times10^{11}M_{\odot}$), the BGGs
present a time sequence where galaxy-galaxy merging increases the BGG
mass by $\gtrsim40$\% ($t_{MGR}\lesssim2$ Gyr) and $V$-band luminosity
by $\sim0.4$ mag.  Our results are remarkably consistent with recent
claims from semi-analytic models that brightest cluster galaxies tend
to assemble at $z<0.5$ \citep{delucia:07b}; such a model also naturally
explains the observed number of brightest cluster galaxies in the
local universe with multiple nuclei
\citep[$\sim20-40$\%;][]{hoessel:85,laine:03}.

Despite the merger activity, the BGGs are dominated by old ($>10^9$
Gyr) stellar populations and show little sign of ongoing or recent
star formation.  The BGGs have rest-frame $(B-V)$ colors corresponding
to mean stellar ages of $>3$ Gyr, and their tight scatter in color
($\sigma_{BV}=0.032$) confirms that the BGGs formed their stars at
$z>0.9$.  If the BGGs continue to age passively, their color scatter
by $z\sim0$ will be even smaller than that of the $\sim100$ brightest
early-type galaxies in Coma ($\sigma_{BV}=0.020$ vs. 0.028).  Their
composite optical spectrum shows no sign of star formation in the last
$\sim1.5$ Gyr.  While BGG-2 \& 3 are weakly detected at \mipsmu, the
X-ray and optical data indicate that BGG-2's \mipsmu~emission is due
to AGN rather than star formation, and possibly for BGG-3 as well.

SG1120 thus provides unique and powerful confirmation of hierarchical
formation on both galaxy and cluster scales because 1) SG1120 has
massive galaxies assembling from galaxy-galaxy merging, and 2) it is a
proto-cluster assembling from the merging of galaxy groups.  Our
findings show that the most massive galaxies in groups and clusters
form via dissipationless merging, and that the galaxy group
environment is critical to forming the massive galaxies dominating
local clusters.


\acknowledgments

K.T. acknowledges support from the Swiss National Science Foundation
(grant PP002-110576) and thanks J. Blakeslee for help during the
initial ACS reduction.  J.M. acknowledges support from
NASA-06-GALEX06-0030 and Spitzer G05-AR-50443, and L.B. from NASA
Spitzer programs through JPL subcontracts \#1255094 and \#1256318.
Support was also provided by NASA HST G0-10499, JPL/Caltech SST
GO-20683, and Chandra GO2-3183X3.

\bibliographystyle{/Users/vy/aastex/apj}
\bibliography{/Users/vy/aastex/tran}

\end{document}